# Addressing the Reality Gap: A Three-Tension Framework for Agentic AI Adoption

Jason Fournier (Imagine Learning)

Kacper Łodzikowski (Adam Mickiewicz University in Poznań)

## Abstract

Generative AI has rapidly entered education through free consumer tools, outpacing the ability of schools and universities to respond. Now a new wave of more autonomous agentic AI systems—with the capacity to plan and act towards goals—promises both greater educational personalization and greater disruption. This chapter argues that successfully navigating these innovations requires balancing three core tensions: (1) Implementation Feasibility, or the practical capacity to integrate AI sustainably into real classrooms; (2) Adaptation Speed, or the mismatch between fast-evolving AI capabilities and the slower pace of educational change; and (3) Mission Alignment, or the need to ensure AI applications uphold educational values such as equity, privacy, and pedagogical integrity. First, we review early evidence of generative and agentic AI in various sectors and in frontline education to illustrate these tensions in context. Then, we present a three-tension framework to guide decision-makers in evaluating and designing AI initiatives across K–12 and higher education. We provide examples of how the framework can be applied to plan responsible AI deployments, and we identify emerging trends—such as curriculum-linked AI agents and educator-informed AI design—along with open research directions. We conclude the chapter with recommendations for educational leaders to proactively engage with the opportunities and challenges of AI, so that this technology can be harnessed to enhance teaching and learning in the decade ahead.

## 1. Introduction: The Generative AI Adoption Shock and the Looming Shift to Agentic AI

In late 2022, the release of the ChatGPT product based on the GPT-3.5 model sent shockwaves through the education sector. Offered as a consumer-oriented application, use of ChatGPT grew at a pace that dwarfed the speed of change in most educational systems. It became the fastest-growing technology on record, reaching about 800 million weekly active users after nearly three years (Bellan 2025). By late June 2025, users were sending roughly 2.63 billion messages per day, of which about 10% were related to tutoring and teaching (Chatterji et al. 2025). With about 46% of those messages coming from users aged 18–25 and intent tilted towards “asking” about half the time, it seems that much of this education use could be driven by students seeking help for tasks such as study planning or tutoring and teachers seeking efficiency through curriculum design or content creation. This inference is strengthened by self-reported data from other contexts. For example, Pearson (2025b) reports over 52% of US college students use generative AI at least weekly for coursework, and 53% prefer AI-powered learning tools to non-AI tools. Among US K–12 teachers, adoption has also surged. In fall 2023, only 18% of K–12 teachers reported using AI for work (Diliberti et al. 2024), yet by the 2024–2025 school year, that number had jumped to 60% (Gallup and Walton Family Foundation 2025).

This rapid consumer transformation resulted in a profound disconnect with the educational system. As of September 2025, only 19% of surveyed higher education institutions have established formal AI policies, and a further 42% were developing guidance (UNESCO 2025). In US K–12 schools, efforts have been similarly uneven. Although 48% of districts reported training teachers in AI by late 2024 (Diliberti et al. 2025), this support has not reached learners. In the first half of 2025, only 35% of district leaders reported providing student training, and over 80% of students said teachers have not explicitly taught them how to use AI (Doss et al. 2025). This ambiguity has fueled anxiety: about half of those students worried about false accusations of cheating with AI, and almost two-thirds of parents feared AI use harms critical-thinking skills (Doss et al. 2025). The AI reality gap between individual and organic adoption and institutional readiness creates both opportunity and risk. Institutions that fail to bridge this gap risk losing relevance, while those that move too quickly without proper frameworks risk compromising educational quality and equity.

While the challenge of generative AI alone was significant, 2025 has ushered in a new wave of agentic AI systems, which can further deepen the consumer–institution divide. In addition to generating a wide range of useful content, they can also act with a degree of goal-oriented autonomy. For example, given the task of buying a course textbook online, an agent first “reasons” about the request, plans the steps, selects a browser tool, navigates to a retailer, compares editions, adds the item to the cart, and completes checkout. This evolves the ability of AI systems from passive question-answering machines towards proactive decision-makers in pursuit of goals. And while questions about the feasibility of scaling such agentic behavior in a safe and secure way remain, the trajectory of the past decade of AI advancements would have us expect that such systems will be robust enough for real-world, at-scale use within a few years. To be ready for the coming shift, we must ask how we move from the promise of agentic AI to practical educational implementations.

In this chapter, we argue that navigating this challenge requires understanding and balancing three fundamental tensions present in any AI-in-education initiative. First, **Implementation Feasibility** asks whether we have the practical capacity and resources to integrate AI meaningfully in classrooms. Second, **Adaptation Speed** highlights the mismatch between the lightning pace of AI innovation and the slower, deliberative tempo of educational change. Finally, **Mission Alignment** demands that AI usage stays true to the educational mission and values, such as equity, privacy, and pedagogical integrity, rather than being seduced by technology for its own sake. All three tensions are at play across K–12 schools and higher education contexts, albeit in different guises and pulling in competing directions.

In the sections that follow, we first lay the theoretical foundation for understanding agentic AI and its potential in education. We then introduce the Three-Tension Framework, describing each tension and its implications. Next, we translate the framework into practical guidance for educational stakeholders to design and evaluate AI initiatives across K–12 and higher education. Finally, we outline our predictions for the immediate future and recommendations for future research. Together, these sections offer a blueprint for turning the promise of agentic AI into sustainable practice in real educational contexts over the next decade.

# 2. From Generative to Agentic AI: The Promise and Early Evidence

## 2.1 Defining Agentic AI

Before proceeding, we must clarify what we mean by agentic AI and how it differs from generative AI and from broader terms such as artificial (general) intelligence. We start from a standard view of AI as the study and construction of agents that perceive their environment and act upon it to achieve goals (Russell and Norvig 2020). This broad definition helps us to appreciate that the ultimate goal of AI has always been the creation of intelligent agents, and that the previous waves of AI (e.g., statistical models of the 1980s, machine learning of the 1990s, neural networks of the 2000s, and generative approaches of the 2020s) were important milestones on that journey. In an educational context, we may employ a complementary policy frame which states that AI systems differ in the types of outputs inferred from inputs (e.g., score prediction, pathway recommendation, content generation), how independently they operate (autonomy), and how they change with use (adaptiveness) (OECD 2024).

Large language models (LLMs) such as those behind ChatGPT fit this AI-system definition but stand out because they can generate fluent outputs across natural languages (e.g., English) and artificial languages (e.g., Python for programming), which is why the label “generative AI” gained traction for these model-centric applications. In practice, generated language is a means to complete downstream tasks such as drafting essays or producing code, so generation is a surface behavior rather than the full extent of system capability.

As language models improved, researchers and developers began to wrap them in structures that allowed them to better interact with the world by planning, calling tools, and maintaining state across steps, bringing the agent idea back to the foreground. An opposing pressure was that—despite the improvements—no single model handled every job well. This situation led to a common computing pattern that assigns narrow, task-specific slices of work to copies of generative models, resulting in a proliferation of task-specific agents. In this approach, agentic AI works in a computing *architecture* that places a model inside a context with typically four main building blocks: a profile that sets role and constraints, a memory that stores and retrieves state, a planning module that structures steps, and an action layer that invokes tools or changes the environment (L. Wang et al. 2025).

In other words, an agent runs a loop: it evaluates input (from the user or another system), plans, acts, and reflects—often using tools and state that persists across turns. This resembles how most users interact with generative systems such as ChatGPT: the user sets context, specifies a task, and expects the model to follow constraints while producing a response—treating the LLM as an external "brain." However, agentic AI differs from this common use in the degree of user control. In standard ChatGPT interactions, the user directs every stage over numerous cycles. By contrast, an AI agent may independently "decide" to seek more information from connected tools or internet-based systems. This autonomy allows the system to change the overall flow of the interaction without immediate human intervention. While the first wave of AI assistants, such as the initial 2022 release of ChatGPT, generated outputs based on their (arguably limited) innate capabilities, an AI agent can choose its own steps and update its working context. In this formulation, the LLM is a component within a broader agent architecture rather than being an agent by itself. This distinction matters for education because policy, safety, and assessment obligations attach not only to the language model offered by a provider such as OpenAI but also to the agency-conferring scaffolding that may be provided by one or more other vendors.

For this chapter, we adopt the predominant mainstream definition of an AI agent that uses tools to accomplish a goal—even if technically that is a redundant definition, given that all of AI is concerned with creating such agents. The distinction between generative and agentic AI matters because the line between them is thin. Fundamentally, both rely on the same large language models. Agentic systems are essentially scaffolded language models that have been fine-tuned to use external tools and to iterate toward objectives. This means they share the innate challenges of language models, such as hallucinations, i.e., confident yet inaccurate outputs. Such issues can be mitigated if the model can use a tool to verify information, but they can also increase if it uses a tool incorrectly to perpetuate an error, such as finding a low-quality reference online and treating it as authoritative.

Finally, we separate agentic capability from artificial general intelligence (AGI) to keep discussion evidence-based. AI providers have floated broad AGI definitions that some people have now conflated with agents, e.g., "highly autonomous systems that outperform humans at most economically valuable work" (OpenAI 2018) or "AI that's at least as capable as humans at most cognitive tasks" (Google DeepMind 2025). A recent measurement-oriented proposal from Hendrycks et al. (2025) defines AGI as a system that matches the cognitive versatility and proficiency of a well-educated adult, operationalized with the Cattell–Horn–Carroll framework into ten broad cognitive components (e.g., knowledge, reading and writing, on-the-spot reasoning, working memory, etc.) and scored as an aggregate "AGI Score" (0–100%). Early measurements of models in that framework (e.g., GPT-4 ≈ 27%; GPT-5 ≈ 57%) show rapid progress over the years, yet a substantial shortfall, with especially weak long-term memory storage and uneven profiles across domains (Hendrycks et al. 2025). In parallel, the OECD's emphasis on gradations of autonomy and adaptiveness rather than a single threshold for "general" intelligence resulted in a recent proposal of AI Capability Indicators (OECD 2025a), which rate systems on five-level scales across nine human-relevant domains (e.g., language; social interaction; problem solving; creativity, etc.). Currently, this framework places leading AI systems in lower-to-mid tiers rather than at human equivalence.

## 2.2 Early Generative AI Implementations Across Industries

Before turning to education, it is useful to review deployments in other sectors, where most investment and large-scale implementation has taken place over the past three years. These implementations present a complex picture: clear gains in bounded, text-heavy workflows, alongside uneven effects and operational hurdles in more complex settings. Large surveys report that roughly seven in ten organizations now use generative AI in at least one function, with concentration in marketing and sales, product and service development, service operations, and software engineering. However, most respondents do not see organization-wide performance gains, suggesting that value remains local to teams unless work is redesigned (McKinsey & Company 2025b).

Perhaps the most illustrative example is software engineering, a domain widely expected to deliver outsized gains because many coding tasks are templated and repetitive. Analysts estimated that AI could help with faster, higher-quality code writing, resulting in a 20–45% uplift in engineering productivity (Chui et al. 2023). Initial reports on narrow slices of work showed positive results, e.g., software engineers using language models completed the task about 56% faster, with a junior developer with about two years' experience saving, on average, roughly 49 minutes more than a developer with about eight years' experience (Peng et al. 2023). However, follow-up studies where engineers worked on real-life software projects showed a mixed picture. For example, in a small-scale randomized controlled trial (RCT), experienced engineers needed 19% more time to complete project tasks than those who did not use language models—once review and clean-up of AI outputs were counted (Becker et al. 2025). This means that, for experienced domain experts working on advanced tasks, any potential AI efficiency gains may be offset by the time required to correct or optimize the AI output.

More consistent benefits are demonstrated in areas with well-documented human processes, such as customer support. In a large-scale field deployment among customer support specialists, access to an LLM-based assistant increased issues resolved per hour by about 15% on average, with the largest gains (around 30%) for less-experienced workers (Brynjolfsson et al. 2025). Here too, however, the effects for the most experienced top performers were smaller and quality sometimes dipped. This pattern was observed in a few other studies on the impact of generative AI on academic and professional productivity. In one RCT, college-educated professionals (e.g., marketers, consultants, data analysts, etc.) doing mid-level writing with AI reduced their completion time by about 37% and raised quality by 0.45 SD, with larger gains for lower-baseline workers (Noy and Zhang 2023). In another RCT, management consultants finished 6–11 minutes faster on an "outside-the-frontier" business case (i.e., analyzing spreadsheet data and interview notes where the right advice required combining numbers with subtle cues from interviews), yet were 13–24 percentage points less accurate (Dell'Acqua et al. 2023).

Taken together, the past three years of generative-AI deployments across industries point to transferable lessons for implementing agentic systems. Organizations that report gains do three things: they redesign workflows around the tools instead of bolting them onto old processes; they pair deployments with clear guardrails for responsible use; and they place oversight with senior leaders who can align investment, risk, and data practice. Where these conditions are absent, post-implementation improvements tend to remain small and local, even when pilot-level results look promising (McKinsey & Company 2025b).

As attention shifts from the reactive generative AI assistants to the more autonomous AI agents, early evidence points to a similar promise and fragility. In one simulation study of expert-validated tasks across sales and service, the best AI agents solved about 58% of single-step tasks but only about 35% of multi-step tasks that required clarification and coordination (Huang et al. 2025). Moreover, the tasks included "confidentiality-awareness" probes that checked whether agents avoided exposing sensitive customer information. All tested models showed near-zero default caution. Mitigating this risky behavior via stricter prompting improved restraint but often reduced task completion. In other simulation studies of everyday internet tasks (e.g., search a site, filter results, open the right page, fill a form, follow multi-step menus, add items to a cart), the best task-success rate ranged between 14% (Zhou et al. 2024) and 23% (Pan et al. 2024). In both cases, common failures included weak website understanding and navigation, poor execution, and limited self-correction. Finally, in a study that moves beyond the browser to full computer use (e.g., email, spreadsheets, coding, media editing), the best agent completed about 12% of tasks against a human benchmark of about 72% (Xie et al. 2024). Errors concentrated in difficulties navigating the graphical user interface and in tool use.

These studies suggest that agents can currently execute short, well-scoped sequences in structured settings, but reliability declines in longer, open-ended interactions with changing interfaces. Moreover, presenting a coherent cross-sector picture for the nascent agentic AI implementations remains challenging because some studies use labels such as "chatbot," "virtual assistant," and "robotic process automation" instead of "agent," which complicates like-for-like comparisons over time and sectors (Kergroach and Héritier 2025).

## 2.3 Frontline Uses of Generative AI in Education

While education was not an early headline use case for generative AI according to initial industry analyses, use grew rapidly and now cuts across three intertwined roles: institutions that adopt AI products and set their rules, staff who integrate tools into teaching and research (often ahead of formal provision), and students who use tools for study (frequently regardless of institutional access or policy).

At the institutional level, adoption follows three main themes familiar from previous waves of educational technology. First, visible pedagogical value: intentions to adopt and reported use rise when generative AI is tied to domain-specific learning outcomes rather than acting as a generic content provider (Alfarwan 2025; Baig and Yadegaridehkordi 2024). Second, capability and confidence: use is higher where teachers grasp the practical basics of AI, can design and review AI-supported tasks, and feel in control of how tools appear in their courses (Lin and Tan 2025; Y. Wang et al. 2025). Third, supportive and safe conditions: adoption is more likely where peers, leaders, and infrastructure back use and where policies and safeguards are clear, with research-informed rules, age-appropriate protections, and reliable access as recurring enablers (Alfarwan 2025; Lin and Tan 2025; Y. Wang et al. 2025). Common barriers include unclear guidance, privacy and data protection concerns, and limited practical support on ethical and assessment practice, reflected in staff reports of weak preparation and strong demand for clearer rules (Baig and Yadegaridehkordi 2024; Lee et al. 2024).

Among staff, one of the biggest use cases is learning design and content creation. Here, usage mirrors the cross-sector pattern of strong gains on bounded tasks but fragile performance on complex work. Teachers and instructional designers report using general-purpose tools such as ChatGPT to brainstorm ideas, draft objectives and explanations, generate quiz items, and streamline related routine tasks (Figueroa de la Fuente and Farhadian 2025; Luo et al. 2025). At the same time, they report risks—e.g., shallow or repetitive tasks, fabricated citations, biased or monocultural content, and blurred authorship and ownership—which make unedited outputs unsafe as final products (Choi et al. 2024; Kozan et al. 2025). It seems, therefore, that generative AI supports education professionals mainly when they retain control as critical reviewers and ethical gatekeepers, systematically checking and adapting AI outputs to local contexts and learner needs. This mirrors the other sectors described above, in which benefits that domain experts derive from AI depend perhaps less on model capability and more on organizational readiness and staff development.

For students, early deployments of off-the-shelf tools such as ChatGPT result in students reporting helpful explanations and feedback across subjects (Ali et al. 2024; Agbo et al. 2025). Structured ChatGPT use has been linked to moderate improvements in learning attitudes and higher-order thinking (Deng et al. 2025). These benefits, however, must be weighed against persistent challenges: the tendency of language models to hallucinate (Ali et al. 2024); the risk that students will outsource their thinking and writing to automated tools rather than using them as scaffolds (Mai et al. 2024); and the unresolved questions of authorship and transparency in assessment systems that rely on take-home, text-based tasks (Jensen et al. 2025). For the latter, early evidence points to constructive use instead of opportunistic misuse. For example, college students using an AI study companion embedded in a courseware platform used it mainly to consolidate core ideas and clarify problems: about 80% of prompts addressed foundational tasks, while around 33% engaged higher-order work such as synthesizing across sources, with some prompts serving both functions (Hendriksen and Lai 2025). Other studies in which student interactions are observed (Ammari et al. 2025) or self-reported (Al-Abdullatif and Alsubaie 2024) show that students use general-purpose AI systems such as ChatGPT for explanations, writing feedback and editing, and high-stakes tasks such as job applications.

## 2.4 Opportunities and Risks of Agents as Active Participants in Education

We can now see that the current wave of agentic AI in education is best treated as the next turn in a long sequence of technological shifts rather than a stand-alone surprise. Agentic AI shifts focus from single on-demand responses to extended sequences of proactive educational decisions. Consider an AI tutor for first-year college students. One early example from the cusp of the generative AI era was Aida Calculus (Pearson 2019), which assessed each line of students' handwritten math against its own generated reference solution and learned to recommend targeted videos and examples through a knowledge graph. It aimed to focus students on method rather than final answers and to provide an always-available companion. Current agentic architectures can take the same idea further. A modern AI math tutor could parse not only students' symbolic expressions but also questions in natural language. It could orchestrate its strategy via specialized agents, such as diagnosis, explanation, practice scheduling, learner analytics, and teacher-intervention alerts. Instead of returning an explanation based on a single interaction, it could query student data across different spaces (e.g., self-study homework and in-class group work) to identify repeated misconceptions and generate targeted follow-up activities adjusted to the learner's interests. The result is a continuous educational intervention based on chains of AI decisions about learner proficiency, content difficulty, timing, and human escalation. These autonomous actions should align with agreed institutional policies and expose decision logs for audit and research.

This vision naturally overlaps with the role of the teacher. Humans have traditionally made all these decisions based on curriculum goals, class constraints, social context at the local and macro level, and professional ethics. Shifting parts of this chain into software raises direct questions about responsibility, transparency, and acceptable autonomy. Institutions will therefore need to treat agents as increasingly autonomous participants in teaching and governance with appropriate guidance and guardrails, not just as better chatbots. This comes at a time when many institutions have barely begun adapting to generative AI. Many districts still grapple with basic questions about grading assistance, citation, and plagiarism detection, while agentic systems introduce more demanding issues. If an AI tutor teaches a subject incorrectly or in misalignment with the language or values of the local community and a student internalizes this as a fossilized error, or if the tutor withholds a meaningful intervention, then responsibility rests with the institutions and vendors that specify, select, and deploy such systems. If agents identify struggling students using patterns that encode inadvertent bias, or if autonomous pedagogical choices conflict with curriculum standards, then schools must retain clear oversight mechanisms and corrective powers. These conditions underscore the opportunity and risk of agentic AI in education. The former lies in truly personalized learning at scale. The latter lies in misplaced trust, opaque decision-making, and over-automation of experiences that require human connection. AI providers and institutions alike need to collaborate on designing capabilities for orchestrating agent behavior, interpreting system recommendations, and maintaining human authority over consequential decisions. Without these capabilities, agentic systems risk becoming unaccountable optimizers that prioritize easily measured proxies such as completion rates while undermining deep understanding, intellectual curiosity, and collaborative capacity. These dilemmas highlight the need for careful balance—an issue addressed in the following sections via our framework of agentic AI adoption tensions.

# 3. How Agentic AI Leads to Three Adoption Tensions

## 3.1 Extending Classic Adoption Frameworks for Agentic AI

The past seven decades of technology development have produced several accounts of how innovation interacts with educational systems. Three perspectives guide this chapter. First, the Diffusion of Innovations (Rogers 2003) frames adoption as a social process shaped by five perceived attributes: relative advantage, compatibility, complexity, trialability, and observability. Rogers' work also distinguishes authority, collective, and individual decision routes. This distinction aligns with how ministries, districts, institutions, and individual educators share power over technology decisions. Second, Ertmer's (1999) framework provides a critical diagnostic for why technology diffusion stalls, identifying two categories of barriers modeled after a practical view of school conditions. First-order barriers are external: dependable access to hardware and software, time, and technical support. Second-order barriers are internal: beliefs about learning, perceived value of technology, and confidence to change practice. Third, effective use ultimately depends on practitioner knowledge, a requirement extended to the digital realm by the Technological Pedagogical Content Knowledge (TPACK) framework (Mishra and Koehler 2006). This model specifies the combined knowledge of content, pedagogy, and technology teachers require to design effective learning.

Generative and agentic AI press against each of these frames. In Rogers' terms, perceived "relative advantage" can be high, yet "compatibility" now extends beyond routines and resources to fundamental pedagogical and ethical values. Educators report concerns about opaque decision-making, data use, and latent bias (McGehee 2024). A system may fit existing workflows and still fail because its autonomy and opacity undermine trust in ways that earlier, passive tools such as word processors or learning management systems did not. Ertmer's barriers, too, take on a different pattern under AI. On the one hand, AI systems can ease some first-order barriers by providing capable, cloud-based tools that reduce on-site software deployment needs. On the other hand, they increase some constraints, such as security and privacy requirements or policy limits. Second-order barriers become even more pronounced. Staff beliefs now include views on acceptable AI use, academic integrity, surveillance risks, and professional identity in AI-mediated classrooms. These dispositions shift slowly and create a persistent trust gap, even where tools appear effective. The classic TPACK model also faces a new challenge. It viewed technology as an instrument under human control. However, AI increasingly behaves as an active partner that suggests, sequences, and sometimes initiates actions. New approaches, such as Intelligent-TPACK (I-TPACK), treat AI as a semi-autonomous collaborator (Chiu 2025). They assume that, because AI operates semi-autonomously, teachers must develop new competencies, e.g., judging AI outputs in alignment with curriculum and values, explaining AI-supported decisions to learners and families, and—if needed—constraining AI actions. The task ahead is therefore not just selecting another tool; it requires rethinking the roles and relationships among teachers, learners, institutions, and the AI systems themselves.

These pressures expose shared limits in classic frameworks. They were developed for technologies that were relatively stable, transparent, and bounded in function, and for institutions that treated adoption as a planned top-down change. Generative and agentic AI instead arrive as fast-evolving, opaque, data-intensive services that raise trust, ethics, and accountability to the foreground. These become practical challenges when they collide with the deep-seated constraints of educational systems, e.g., funding rules, procurement cycles, or regulatory obligations. The next section turns to the structural conditions of K–12 and higher education that shape AI adoption choices and sets up the three tensions that will frame our analysis.

## 3.2 Systemic Constraints on Agentic AI in Education

### 3.2.1 K–12: Law, Equity, and Risk Management

The application of the above-mentioned frameworks inevitably intersects with the reality of today's educational systems. K–12 schooling, in particular, is designed to ensure consistency of provision under strong legal, political, and public scrutiny. That design favors predictability and procedural safety over rapid change, which slows the adoption of complex technologies such as generative and agentic AI.

To start with, the policy landscape is a complex mix of controls such as federal and state laws, academic standards, and accountability measures that AI systems designers must respect. Policy variation across states means there is no single reference curriculum. For example, California mandated ethnic studies as a graduation requirement for the class of 2030 (California Department of Education 2021), while Florida restricted the instruction on race and gender (Schoorman and Gatens 2025), and Texas removed the requirements to teach about civil rights movements (Texas Legislature 2021). This means that an AI system that appears aligned in one context may be unacceptable in another.

Second, the student population is diverse and unequal. In the US, about 11% of students are English Learners (National Center for Education Statistics 2024a) and 15% receive disability services (National Center for Education Statistics 2024c), while Black and Hispanic students are more likely than White students to attend high-poverty schools (National Center for Education Statistics 2024b). These distributions matter for AI design. Most leading models perform best in high-resource languages, handle some dialects and disability-related needs less reliably, and generally draw on training data that disproportionately favors majority perspectives. If schools deploy AI without adjustment and monitoring, they risk deepening racial, linguistic, and disability-based gaps rather than reducing them. Another consideration is access to AI. Students in higher-income families are more likely to have personal devices and paid AI licences. Restrictive or vague school policies can therefore widen gaps in effective use. District capacity also differs in ways that can compound these gaps. In fall 2024, 67% of low-poverty districts reported training teachers on generative AI, compared with 39% of high-poverty districts (Diliberti et al. 2025).

Third, data protection and accessibility rules create hard constraints. Data protection laws such as FERPA set the conditions for when schools and vendors may collect, use, and share student information, and require clear limits, security safeguards, and transparency (Family Educational Rights and Privacy Act 1974). Tools that handle children's data face additional restrictions that limit tracking, profiling, and commercial reuse, and push districts to negotiate contracts that prevent vendors from using student content to train general models (Children's Online Privacy Protection Act 1998). In parallel, recent accessibility rules require public schools to align websites, learning platforms, and mobile applications with Web Content Accessibility Guidelines (WCAG) on a defined timetable (US Department of Justice, Civil Rights Division 2024). Many AI products do not yet meet these privacy and accessibility expectations as delivered, so districts run detailed legal, security, and accessibility reviews before adoption, which further slows deployment.

Finally, the budget landscape adds another layer of complexity. District budgets draw on a mix of federal Title programs, state allocations, and local property taxes, which provide about 80% of local K–12 education funding (Lincoln Institute of Land Policy 2022). The expiration of nearly $200 billion in Elementary and Secondary School Emergency Relief (ESSER) funds has left districts with significant structural pressures as they try to sustain pandemic-era staffing, support, and technology (Center on Budget and Policy Priorities 2024). Within this constrained context, Title I funds (intended to ensure equitable access for students from low-income families) sit inside strict spending rules: they must add to, not replace, existing support; districts must keep funding at a stable baseline; schools must receive fair treatment across the district; and all uses must be evidence-based and tied to documented need (US Department of Education 2019). Districts must then move prospective AI purchases through multi-stage vetting for curriculum alignment, data protection, accessibility, and contract compliance. The resulting procurement cycles often span six to eleven months, with 25% of leaders reporting timelines exceeding a year for core curriculum purchases (Fittes 2025). AI capabilities move on a quarterly rhythm, which creates a fundamental mismatch between the pace of technological innovation and the deliberate, stakeholder-intensive process of educational change.

### 3.2.2 Higher Education: Fragmented Authority and Uneven AI Use

Higher education operates under different constraints but faces similar challenges. Universities must balance academic freedom with institutional coordination, navigate shared governance structures that distribute decision-making across faculty senates and administrative hierarchies, and manage the competing demands of research, teaching, and community engagement. Calls for shared responsibility among governing boards, administrations, and faculties have existed for decades (e.g., American Association of University Professors 1966). Yet, evidence of effectiveness remains "sporadic at best," and in practice, these shared structures often yield limited influence over major decisions (Miller and Nadler 2019). The tension between individual academic freedom and institutional coordination requirements has intensified as educational technology, including AI, becomes more integrated with minimal oversight (AAUP Committee on Artificial Intelligence and Academic Professions 2024).

The decentralized nature of academic departments means AI adoption often happens in silos. In one study, interviews revealed concerns that decentralized decision-making makes coordination difficult, with different programs developing separate policies and resulting in a lack of standardization at the institutional level (Baytas and Ruediger 2025). Computer science departments may implement in-house AI tutors for coding (Liu et al. 2024), while humanities departments may have philosophical objections to the use of AI or lack the training and infrastructure required for implementation. This is reflected in early analyses of the usage of AI systems. In one study of a million student conversations, computer science students generated around 37% of conversations while comprising only around 5% of US degrees, dramatically overrepresenting their share of adoption, while humanities students were significantly underrepresented (Handa et al. 2025).

#### 3.2.3 Institutional Readiness: The Missing Infrastructure Nobody Planned For

Educational institutions have worked hard to build their technological foundations, but that work took significant effort, and those foundations likely cannot change quickly. Federal funding covers network infrastructure and cloud-based virtualized equivalents of eligible internal connections equipment, but not general cloud computing services unrelated to connectivity infrastructure (Federal Communications Commission 2019; Eggerton 2019). State education budgets, already strained by teacher shortages and facilities maintenance, lack line items for AI operations. As of December 2021, public K–12 employment was down 376,300 positions (4.7%) from pre-pandemic levels (García and Weiss 2022), while schools face an $85 billion annual shortfall (rising to at least $98.6 billion with inflation) in funding for proper maintenance and upgrades, with the average school building reaching 50 years of age (K–12 Dive 2024; US Government Accountability Office 2020). It is in this social-historical context that the current generation of AI tools emerged, leaving schools to face an uncomfortable choice: deploy limited pilots that demonstrate possibility but lack transformative scale, or commit resources they do not have to systemic implementations that might fail.

Professional development presents another bottleneck entirely. Deployment of agentic AI will likely require training on AI literacy, training on specific systems, and a thorough reassessment of pedagogical practices. Teachers must understand not just which buttons to click but how agents make decisions, when to intervene in autonomous processes, and how to interpret algorithmic recommendations within broader pedagogical contexts. A recent study of teacher preparation programs found that while 59% provide some AI-related instruction, only 25% provide training on ways AI can support teaching, with most instruction focusing on plagiarism prevention rather than AI literacy or integration (Weiner et al. 2024). Among US teachers who had not used AI in teaching in the previous 12 months, 70% felt they lacked the skills to teach with AI, and 27% cited insufficient school infrastructure (OECD 2025b). Mirroring the adoption of AI in the broader workforce, participants in the education process may require retraining, upskilling, or reimagining their roles in the face of advancing AI technology.

### 3.3 From Constraints to Adoption Tensions

Classic adoption frameworks still guide AI decisions in education, yet agentic AI and current system conditions push beyond what those models were built to handle. They assumed bounded tools, gradual change, and relatively clear decision routes. Agentic AI, however, arrives as a fast-moving technology into an educational world marked by multiple veto points, tight compliance duties, and uneven access and capacity.

Together, these conditions compress risk and choice into three recurring tensions that shape agentic AI adoption: whether implementation is practically feasible, whether institutional adaptation can keep pace with technological change, and whether uses remain aligned with educational mission and equity. The next section develops these tensions into a structured framework for analyzing decisions and designing responsible implementations.

## 4. The Three-Tension Framework

In this section, we formalize the three tensions that arise when agentic AI meets current educational systems and use them as a practical guide for educators, researchers, and institutional leaders. Our framework asks three main questions:

- **Implementation Feasibility:** Can institutions run a given AI system safely and sustainably, beyond a pilot, with their real infrastructure, staffing, funding, and legal obligations?
- **Adaptation Speed:** Can institutional learning, decision-making, and policy cycles keep pace with rapid AI capability and product change?
- **Mission Alignment:** Do proposed uses of AI support educational purpose, equity, trust, and learner agency, rather than pulling practice towards shallow efficiency?

Table 1 summarizes these tensions and the types of responses they call for.

Table 1. Overview of the Three-Tension Framework for Agentic AI Adoption

| Tension | Core question | Root causes | If mishandled | Productive response |
|---|---|---|---|---|
| Implementation Feasibility | Can we run this system safely and sustainably beyond a pilot? | Funding limits; aging infrastructure; data protection and accessibility obligations; vendor dependence; limited technical and legal capacity; thin professional development. | High-profile pilot failures; security or privacy incidents; unsupportable contracts; tools abandoned after grant cycles; staff fatigue and distrust. | Start from clear use cases and outcomes; cost and risk models; staged pilots with exit criteria; governance, monitoring, and human-in-the-loop protocols; honest readiness checks before scale-up. |
| Adaptation Speed | Can our decision and learning cycles keep pace with AI change? | Slow budget and procurement cycles; academic calendars; multi-layer governance; uneven AI literacy; rapid vendor release cycles and marketing pressure. | Policy vacuums that fuel concealed or shadow use; reactive bans; fragmented experiments; waste on short-lived tools; inability to learn from trials. | Use rolling roadmaps; separate sandbox from core systems; gate progress through staged adoption; invest in shared learning capacity so the system can absorb change. |
| Mission Alignment | Does this use of AI serve our educational purpose and equity goals? | Assessment models that reward fast outputs; pressure for efficiency; opaque machine learning models; biased data; unequal access to devices and AI tools. | AI-as-shortcut culture; hidden bias against minoritized learners; surveillance uses that erode trust; two-tier provision (human-rich for some, AI-heavy for others). | Redesign assessment to surface students' own thinking; apply equity and ethics checks; keep humans in charge of consequential decisions; align AI roles with clear values on learning, care, and civic preparation. |

These tensions form intersecting continuums. Every design choice for agentic AI moves an institution along all three at once. A cautious, narrow pilot may protect Mission Alignment but expose weak Adaptation Speed and brittle Feasibility when scaling is attempted. A rapid, well-funded rollout can look strong on Feasibility and Speed while undermining trust, equity, or assessment integrity. Blanket restrictions intended to defend values can instead push activity into concealed or shadow use, weakening safety, data protection, and institutional learning.

Our framework provides a diagnostic way to see where institutions sit within these tensions and practical strategies for navigating them—for these are tensions to be managed, not problems to be solved. Each represents a continuum where tilting too far in one direction can jeopardize success. The framework helps institutions locate where pressure concentrates, see where they may be over-correcting, and select responses that manage trade-offs consciously across the set.

We propose the framework as an extension, not a replacement, of established models. Implementation Feasibility grounds Ertmer's first-order barriers (access, time, and support) in real constraints: procurement cycles, professional development capacity, and security requirements. Adaptation Speed maps Rogers' diffusion dynamics onto institutional rhythms, helping leaders pace trials and scale up across authority, collective, and individual decision routes. Mission Alignment builds upon TPACK, extending beyond technical competence to include the foundational judgment of educators about the role of AI in education as well as the role of education in the AI-powered job economy. While earlier literature provides the conceptual map, our three tensions mark the junctions where agentic AI forces explicit trade-offs. The next subsections examine each tension in detail.

### 4.1 Implementation Feasibility: The Gap Between Technical Pilots and Sustainable Solutions

Across industries, 79% of organizations report using AI agents in some capacity (PwC 2025), yet over 40% of agentic AI projects will be canceled by 2027 due to escalating costs, unclear business value, or inadequate risk controls (Gartner 2025). In education, generic tools such as ChatGPT have seen widespread individual uptake, but more specialized, domain-specific systems are only starting to reach full implementation and evaluation. Common challenges include value propositions that appear attractive in small trials but cannot justify costs at scale, technical brittleness that surfaces under real-world load, inadequate monitoring and support infrastructure, weak or fragmented data that limits AI performance, and underestimation of the organizational change required. Implementation feasibility therefore extends beyond whether an institution can pay for licenses. It asks whether the institution has the financial, technical, legal, and human capacity to integrate AI safely and sustainably into everyday practice.

A recent case from K–12 education illustrates this feasibility gap. In early 2024, Los Angeles Unified School District launched an AI-powered chatbot called “Ed” as part of a high-profile $6 million initiative. Within three months, the project collapsed amid allegations of misused student data, unrealistic timelines, and the vendor’s inability to deliver a reliable product (Chapman 2024). This stands in contrast to Georgia Tech’s “Jill Watson” assistant, which began in one course with a narrowly defined support role, was refined iteratively over several years, built internal expertise and trust, and only then expanded its scope once its value had been demonstrated (Martin 2025). The contrast between “Ed” and “Jill Watson” shows that feasibility rises when scope is narrow, goals are explicit, and pilots are treated as structured learning investments with evidence-based decisions about scale.

First, there must be a clear value proposition. Educational leaders should be able to state which learners or staff the AI system will support, which outcomes it is expected to change, and through which mechanism. Vague promises of “personalization” or “efficiency” are insufficient. A defensible case might specify, for instance, improved feedback turnaround in large first-year modules or a reduction in course withdrawal for identified risk groups, and then require that pilots generate evidence against these claims before any expansion.

Second, expectations about return on investment must match the realities of institutional change. While many AI pilots show promise, they may not initially show rigorous evidence of efficacy at scale or consistent performance returns (Giannakos et al. 2024). Meaningful benefits may take 12–18 months or more to appear, whereas many institutions hope for results within a single term. Feasible plans treat pilots as opportunities to gather data and refine design, not as guarantees of immediate gain. They include explicit success criteria, timelines, and exit conditions, and they recognize professional development as a funded component rather than assuming educators will absorb new tools and practices without support.

Third, production use demands robust safety, monitoring, and equity mechanisms. Manual checking of AI outputs is viable for a small pilot but not for thousands of daily interactions. Scalable systems need logging, automated alerts, and human-in-the-loop protocols to identify harmful or inappropriate content, signs of student distress, academic integrity risks, and technical failure modes. They should also support disaggregated analysis so that institutions can see how performance and error patterns vary for multilingual learners, students from different racial and ethnic groups, students with disabilities, and other equity-relevant cohorts. Without such visibility, feasibility claims ignore the risk of uneven harm once deployment reaches diverse classrooms.

Fourth, data privacy, security, and regulatory alignment must be resolved before scale-up. Institutions should conduct formal data protection impact assessments and ensure that contracts specify how student data is collected, processed, retained, and deleted, consistent with laws such as the FERPA. "Early access" offers or pilot schemes must not bypass standard safeguards. Vendors that request unusually broad rights to reuse student data, or cannot explain their security and retention practices clearly, present clear warning signs. In higher-risk scenarios, regulatory sandboxes can provide a controlled path: time-boxed pilots under defined safeguards and independent scrutiny, designed to surface risks and evidence before wider deployment (Gasser and Schmitt 2023).

Fifth, institutions need sufficient infrastructure and policy readiness. Many frameworks highlight conditions such as a shared vision, secure and reliable infrastructure, clear policies, equitable access, and sustained professional learning (EDUCAUSE 2024; CoSN and CGCS 2024; ISTE 2025). Readiness assessments based on these resources help identify whether leadership, governance, infrastructure, staff capability, and funding can support AI at scale. Where diagnostics indicate weaknesses, institutions should address these gaps before committing to large-scale deployment.

Taken together, these considerations act as decision gates for Implementation Feasibility. Many well-intentioned AI projects falter precisely at the transition from grant-funded or temporary pilots to sustained programs. If an AI initiative does not demonstrate clear value, realistic timelines and costs, reliable safety and equity monitoring, strong privacy and security safeguards, and institutional readiness, then it should remain a bounded experiment rather than move into system-wide production.

### 4.2 Adaptation Speed: Innovation Cycles Versus Institutional Change

The second tension, Adaptation Speed, captures the gap between how quickly AI capabilities advance and how slowly educational systems can change rules, skills, and infrastructure. Frontier AI labs iteratively evolve language models on a cycle of months. By contrast, curriculum reforms, procurement, accreditation, and shared governance often move on three- to seven-year timelines. Recent examples underline this mismatch. OpenAI released GPT-4 in early 2023 (OpenAI 2023) and introduced more advanced "reasoning" models by late 2024 (OpenAI 2024). Over the same period, many schools and universities were still drafting their first policies on chatbots. In a 2024–25 survey, 68% of teachers reported no training on AI, despite growing expectations that they would guide students' use of these tools (Gallup and Walton Family Foundation 2025). By the time institutions respond to one wave of tools, the technology has already shifted.

Institutions tend to respond to this moving target in two suboptimal ways. Some wait for AI to "settle" before acting. This delay creates a vacuum that staff and students fill with unsanctioned use of consumer tools. One cross-sector study found that 56% of employees used AI at work without clear approval and nearly half had shared sensitive data with public systems (Gillespie et al. 2025). Schools that block or ignore AI face similar unapproved use, only with weaker safeguards and no shared learning. Others rush in the opposite direction, adopting multiple AI products quickly to appear innovative. Without time for alignment, training, or integration, many of these tools see low uptake, duplicate existing systems, or erode trust when they fail. Early education surveys reflect this divide: about one-third of teachers reported using AI weekly and saving several hours per week, while roughly 40% did not use AI at all (Gallup and Walton Family Foundation 2025). The gains concentrate where support exists; elsewhere, potential benefits remain unrealized.

Past technology waves suggest that educational change has a stable tempo. Computers, internet connectivity, and learning management systems all followed decade-scale adoption curves from early use to ubiquity. Moving from pilot to meaningful pedagogical integration typically took longer still. Generative and agentic AI may spread faster than previous tools, but institutional processes will not compress to match quarterly release cycles. Meanwhile, commercial incentives favor rapid feature launches over the less visible work of integration, support, and evaluation. The result is a structural tension: AI changes faster than education can, yet education cannot opt out.

Managing Adaptation Speed well does not mean being first to adopt every new system. It means building the capacity to learn, adjust, and scale in a predictable way. First, institutions can use rolling roadmaps. A roadmap sets priorities over one to two years—such as feedback, tutoring, analytics, or support for specific learner groups—and is reviewed on a fixed schedule, e.g., each quarter. This approach gives stability for planning and professional development, while creating formal points to adjust in response to new evidence, capabilities, or regulations. Clear ownership of the roadmap ensures that changes are deliberate rather than reactive.

Second, leaders should separate pilots from core operations. Sandboxes such as summer courses, elective modules, after-school programs, or voluntary faculty communities allow experimentation without placing high-stakes decisions, grading, or credentialing at risk. For each pilot, institutions should state in advance the conditions under which it will be expanded, redesigned, or stopped, including evidence on learning outcomes, safety, usability, and staff support.

Third, staged adoption with explicit gates can align pace with capacity. An initial stage limits use to small groups with close monitoring. A second stage widens use once early criteria are met and includes targeted training and policy guidance. A third stage moves to broader deployment only after the system proves reliable, aligned with mission, and manageable within available resources. Each transition requires a conscious decision based on data, not momentum.

Fourth, policies should rest on cross-cutting principles rather than narrow assumptions about interface. AI will not remain a text-only chatbot. It is already appearing as embedded document assistants, voice agents, learning platform copilots, and multimodal tutors. Institutions can reduce policy churn by focusing on stable requirements—such as transparency about when students interact with AI, documentation for AI-supported work, human review of high-stakes decisions, and consistent privacy protections—so that rules remain applicable as interaction modes change.

Fifth, institutions need to invest in absorptive capacity: the ability to notice relevant developments, interpret them, and apply them effectively (Cohen and Levinthal 1990). This includes structured professional learning on AI, time for educators to share trials and results, cross-functional teams that can read technical and legal guidance, and norms that treat small, well-governed experiments as expected practice. Emerging survey evidence suggests that schools with explicit AI policies and support report larger productivity gains from the same tools than those without such structures (Gallup and Walton Family Foundation 2025). The difference lies less in access to technology and more in the capacity to adapt.

Adaptation speed, understood this way, is not a race to adopt faster than others. It is an institutional skill: to respond to rapid AI change without defaulting to paralysis or hype, and to ensure that each step strengthens, rather than destabilizes, educational practice.

### 4.3 Mission Alignment: Educational Values in an AI-Mediated World

The third tension, Mission Alignment, asks whether AI in education supports the core purposes central to an educational institution, which may include domain understanding, critical thinking, creativity, and preparation for civic and economic life. Some commercial AI systems, however, are built to optimize engagement, throughput, or short-term performance metrics. If adopted on those terms, they can start to steer teaching, assessment, and support toward what systems optimize rather than what educators value. Survey data reflects this unease: many teachers report that AI can improve the timeliness and personalization of feedback, yet a majority also fear that heavy reliance on AI will weaken students' capacity for independent thinking and original work (Gallup and Walton Family Foundation 2025). Mission alignment is therefore about using AI to support learning, not allowing AI to redefine what counts as learning.

Mission alignment is especially visible in four areas: assessment, equity, human roles and agency, and preparation for AI-shaped work and life, with emerging frameworks providing criteria for practice. First, assessment is the most immediate test of whether AI use reinforces or erodes educational aims. When high-stakes tasks reward polished products, rote recall, or formulaic writing, students have strong incentives to use AI as an authoring shortcut. Aligning AI with educational aims instead requires tasks that make thinking visible. Research on active learning shows that students learn more when they generate, discuss, and apply ideas than when they simply receive information (Freeman et al. 2014). Generative AI can fit this model if assignments require students to explain, critique, and adapt AI outputs rather than submit them unchanged. As teaching and learning move into a dense "digital ocean" of learner activity, this shift calls for movement from episodic, decontextualized tests toward continuous interpretation of authentic digital traces (DiCerbo and Behrens 2014). Work on complex digital performances shows how rich tasks and simulations can shorten the inferential distance between observed behavior and valued competencies, providing a more defensible basis for high-stakes decisions (Behrens et al. 2019). In this view, AI environments are not just answer engines; they are spaces where ongoing human–AI interaction can support learning and, if assessment is aligned accordingly, yield trustworthy evidence of what learners know and can do (Łodzikowski et al. 2024). Well-designed assessments make appropriate AI use and genuine thinking coincide: students earn credit for how they interpret, adapt, and justify AI-supported work.

Equity is a second critical dimension. Without deliberate attention, AI access and quality can track or exacerbate existing inequalities. Students with reliable devices, connectivity, and private access to advanced tools are better placed to benefit, especially if schools formally restrict AI but cannot prevent informal use. In such settings, bans tend to advantage those with more resources and disadvantage those without. A mission-aligned response provides structured access to AI for all students and integrates AI literacy into the curriculum so that effective and responsible use is not limited to a privileged subset. At the same time, AI systems themselves can encode bias. Studies report, for example, that some language technologies misjudge or devalue writing by non-native English speakers, and that speech and image recognition systems show higher error rates for people of color and people with disabilities (Liang et al. 2023; Koenecke et al. 2020). Institutions, therefore, need to test candidate tools on representative data, monitor outcomes with disaggregated metrics, and keep humans in charge of high-stakes decisions. When AI influences judgments about risk, placement, or access to opportunities, educators must have both visibility into system behavior and authority to correct biased patterns.

Human roles and agency form a third area. Education depends on relationships, trust, modeling, and professional judgment. Mission alignment requires AI systems that extend this work rather than replace or obscure it. In practice, AI should take on clearly defined support tasks—such as drafting practice items, suggesting explanations, or highlighting patterns in student performance—while teachers, advisors, and leaders remain responsible for interpretation, ethical decisions, and care. Early-warning and analytics systems are a concrete example: predictive models may flag students at risk, but humans decide how to respond, talk with students, and tailor support. One high-level guideline is for humans to set goals and limits and for AI to assist with analysis and scale, ensuring that humans retain control over consequential decisions. Experience from other fields supports this stance. Predictions that deep learning would rapidly make radiologists obsolete (Mukherjee 2017) sit alongside subsequent evidence of persistent workforce shortages and widespread use of AI as an assistive tool, not a replacement (Mousa 2025). AI took over narrow perceptual tasks, but radiologists' roles in explanation, integration, and accountability remained central (Krishna Nk et al. 2025). There is a direct analogy for education; as AI automates parts of content design and proficiency assessment, the distinctively human aspects of teaching—motivating students, contextualizing knowledge, guiding values, sustaining communities—become more, not less, important. Mission-aligned design uses AI to create more space for that work.

Preparation for future work and life is the fourth area. There is no consensus on whether LLMs will mainly complement or displace white-collar jobs. Some analyses suggest that capabilities such as complex problem solving, creativity, and ethical reasoning attract wage premiums when combined with AI (Mäkelä and Stephany 2024). Others suggest that AI may automate parts of high-skill jobs and compress wage distributions by reducing some top-end earnings (Bloom et al. 2024). Using a rubric that evaluates whether LLMs can reduce the time required for a task by at least 50% with no loss in quality, Eloundou et al. (2024) estimated that roughly 80% of the US workforce has at least 10% of its tasks exposed to this acceleration, while 19% has more than 50% exposed. They also found that LLMs alone could accelerate about 15% of tasks, while LLM-powered applications increased this share to as much as 56%, extending potential impacts into many higher-wage professions. As of late 2025, however, US labor data does not show an abrupt AI-driven employment shock; occupational change since 2022 appears similar to earlier periods, and many "AI layoffs" seem tied to broader restructuring, with AI cited as one factor among several (Gimbel et al. 2025). A counterpoint is that traditional economic indicators such as census data or GDP can miss the depth of AI transition because they measure outcomes after disruption rather than technical exposure, where AI capabilities overlap with human skills. By mapping more than 13,000 AI tools to job skills, Chopra et al. (2025) argued that the most visible adoption of AI in the tech sector accounts for just about 2% of wage value ($211 billion), while about 12% ($1.2 trillion) lies hidden in administrative and financial services. Yet theoretical exposure does not equal imminent agentification. In benchmarks of real-world freelance work from platforms such as Upwork, current-generation agents perform near the floor on end-to-end project execution, with the leading agent completing only 2.5% of commissioned projects to an acceptable, human-evaluated standard (Mazeika et al. 2025). In this uncertain landscape, a mission-aligned stance does not chase every transient technical skill of current tools. It emphasizes durable capacities—critical thinking, learning how to learn, collaboration, communication, and ethics—that enable learners to adapt across changing technologies and roles.

These commitments are reinforced by existing guidance frameworks that translate abstract values into sector-specific standards. On a global policy level, UNESCO (2021) emphasizes human agency, inclusion, equity, and the public interest. Within higher education, EDUCAUSE (2025) codifies these into ethical principles including beneficence, justice, and autonomy.

Similarly, TeachAI (2025) adapts these concerns for the K–12 context, stressing purpose-driven use, compliance, and student agency. Collectively, these frameworks provide concrete criteria for evaluating AI proposals: not only whether a system functions or is affordable, but whether it strengthens learning, protects rights, supports fairness, and preserves meaningful human roles.

Within the three-tension framework, Mission Alignment is not an afterthought once feasibility and speed are addressed. It is the primary filter for deciding which AI uses are worth pursuing and how implementation choices should be judged over time.

### 4.4 Navigating Tension Interactions

The three tensions described above—Implementation Feasibility, Adaptation Speed, and Mission Alignment—do not operate in isolation. They interact and compound. Action on one dimension almost always shifts pressure onto the others. Most serious AI initiatives will push on all three at once, and trying to fix one in isolation often creates new problems in the others.

One common pattern is defensive over-correction in the name of integrity. A school may seek to protect academic standards by banning AI in most assignments. That stance appears to support Mission Alignment, yet it can push AI use into unsupervised spaces, prevent staff from building capability, and block systematic learning about how to manage the tools. In this way, strict rules weaken Adaptation Speed and lead to unfeasible, opaque practices beyond institutional oversight. An opposite pattern arises when a forward-looking university pursues a rapid—and possibly high-profile—AI rollout. This may address Adaptation Speed, but if IT support, data protection, accessibility, and professional development lag behind, the feasibility may unravel and staff or students may judge the deployment as misaligned with core educational values.

Overall, these patterns are sharpest at the step from small pilots to sustained programs. A pilot can align well with institutional values and teaching aims (Mission Alignment), yet attempts to scale may reveal unresolved costs, infrastructure gaps, privacy risks, and support needs (Implementation Feasibility). If leaders rush from pilot to broad deployment to appear responsive to fast-moving technology (Adaptation Speed), any visible failures may erode trust and prompt restrictive responses that undermine long-term goals. In practice, most apparent “AI failures” in education arise less from limitations of the underlying models and more from suboptimal management of the three tensions.

Managing these linked tensions requires coordinated action, not isolated fixes. Technical capacity must grow in step with human and organizational capacity. Buying an AI system without time and support for instructors and advisors leaves even strong tools poorly used. Policy and governance need iterative updates: provisional guidelines, clear review processes, and scheduled revisions reduce both the vacuum that leads to unsanctioned use and rigid rules that age badly. At each decision point, leaders can apply a small set of mission-focused questions: How does this use support students’ learning and well-being? How does it affect equity? What evidence will we use to judge impact? Which trade-offs remain unacceptable even if the system performs technically well?

In this sense, adopting agentic AI is a continuing process, not a single decision. Institutions that keep feasibility, speed, and mission in view at the same time—checking capacity, pacing change, and holding AI uses to clear educational purposes—are more likely to reach stable, defensible practice. Institutions that treat only one tension as primary risk bans that drive use underground, rollouts that collapse, or innovations that erode core aims. Making the three tensions explicit, and managing them together, turns them into a practical guide for decision-making rather than hidden sources of failure.

# 5. Emerging Trends and Future Research Directions

### 5.1 Curriculum-Grounded, Platform-Bounded Agents

A visible commercial pattern is the move from open, general-purpose chatbots to agents embedded in existing educational platforms. Providers now build agents into courseware, textbooks, assessments, and learning management systems. These agents draw on vetted content aligned to standards, preferred theory of learning, and institution-owned data rather than import hidden assumptions baked into general-purpose solutions (Imagine Learning 2025; Pearson 2025a).

In this model, institutions can constrain agents to approved content libraries, policies, and tools. Agents retrieve tagged resources that express learning goals, standards, misconceptions, and success criteria. They log which sources inform substantive outputs. They call only those tools that meet privacy, accessibility, and safety requirements. When a request falls outside this governed space or depends on external systems, default behavior is to decline, refer, or escalate.

This pattern offers a practical near-term route to agentic AI. It supports Implementation Feasibility by reusing platforms and reducing bespoke build. It supports Adaptation Speed by allowing incremental changes inside known systems. It supports Mission Alignment by tying agent behavior to local curriculum, assessment norms, and values, rather than opaque external content.

### 5.2 Multi-Agent Networks Across Products and Institutions

Outside education, industry roadmaps now assume a shift from single assistants toward connected “agentic workforces” operating across systems. Reports on OpenAI’s five-level roadmap outline a progression from chatbots (Level 1) and problem-solving “reasoners” (Level 2) through tool-using agents (Level 3) toward innovative systems (Level 4) and, at the horizon, organization-level systems able to manage complex workflows (Level 5), with the upper levels presented as directional rather than current capability (Cook 2024). Management consulting firms describe a similar future, in which companies build an “agentic organization”: a managed network of human and AI agents coordinating work across applications and data rather than a single assistant bolted onto existing processes (McKinsey & Company 2025a).

In education, this trajectory could translate into a mesh of role-specific agents that work across products and institutions under clear policy, identity, and audit constraints. A learner's personal agent could authenticate to handle internal tasks, such as reading course requirements and invoking approved tutoring, library, or accessibility services. With explicit consent, it could then bridge to such external sources as consulting labor market feeds and credential frameworks to propose end-to-end study-to-work pathways. Program- and system-level agents could draw on partitioned signals from courses and campuses to track engagement, surface inequities, and anticipate support needs without exposing full individual records, while external agents interact with institutional agents only through protocols that define who may request what, how data is minimized, and how each significant action is logged. In such a design, institutions do not cede control to opaque commercial assistants: they specify which agents may interact, set guardrails once at the network level, and keep grading, progression, placement, and welfare decisions with accountable humans. This networked model scales by reusing shared components instead of multiplying disconnected tools, can adapt as capabilities change without constant rebuilds, and keeps educational purpose, trust, and equity encoded in the rules that govern how agents work together.

### 5.3 More Opportunities for Educators to Co-Design

Another emerging trend is the ability for educators to turn general-purpose language models into local, task-specific tools with little or no code. Instructors can already prompt or lightly scaffold models to act as course-aligned tutors, generate visualizations or simulations for specific concepts, script practice dialogues, or build simple games and study companions that reflect their own materials and expectations (Bent et al. 2025). The same underlying capability lets instructional designers experiment quickly: they can prototype an agent that walks students through a problem-solving routine, enforces "think-before-answer" steps, or provides feedback in a defined tone. They can then refine that behavior based on classroom evidence rather than vendor assumptions.

Additionally, educational platforms are beginning to expose configuration layers that let institutions and educators shape agentic behavior directly. Instead of accepting a fixed "AI assistant," faculty and program leads can adjust scoring rubrics, define allowable sources, set escalation rules, or constrain how agents handle citations, style, and academic integrity, often through policy panels or structured natural-language instructions. Early moves in this direction treat educators less as tool recipients and more as partners who specify how tutoring, grading support, or advising agents should behave inside their context—including at the stage of designing the product (Hardman 2025).

Framing educators as co-designers in these ways strengthens all three tensions in our framework. It improves Implementation Feasibility because agents are built around real curricula, policies, and workflows. It improves Adaptation Speed because those closest to learners can iterate within shared guardrails instead of waiting for distant release cycles. It supports Mission Alignment because norms about transparency, effort, equity, and care are encoded in the agent's prompts, constraints, and hand-off rules, keeping human judgment central while using agentic systems as extensible, educator-shaped infrastructure.

### 5.4 Research Gaps

Despite the fast-growing interest in AI for education, several fundamental research gaps remain. First, developing robust benchmarks and shared datasets is an urgent priority. There is a need for standardized tasks and evaluation frameworks to assess agentic AI systems on educational outcomes, safety, and equity. Many claims about AI "passing" exams or outperforming humans lack validation in real classroom contexts. Researchers should create education-specific benchmarks—such as complex multi-step problem solving or pedagogy-aware tutoring scenarios—and open datasets drawn from authentic student work. These resources would enable consistent comparison of AI systems and toolkits for improving them. They would also encourage researchers and vendors to demonstrate results on transparent, agreed-upon metrics rather than cherry-picked examples. In tandem, more rigorous pre-deployment testing is needed: before schools adopt AI tools, they should be evaluated on curriculum-aligned tasks and stress-tested for reliability, bias, and accessibility issues. Building this evaluation infrastructure is a practical step to ground AI promises in evidence.

Second, more studies are needed on how students and teachers interact with AI and how these interactions influence learning. For example, how does extensive reliance on AI tutors or writing aids affect a student's development of problem-solving skills, critical thinking, or confidence in their own abilities? Does using AI for immediate answers discourage persistence, or can it be harnessed to actually strengthen metacognitive skills? Longitudinal and mixed-method research should examine such questions across diverse age groups and subjects. This includes classroom observations, think-aloud protocols as students work with AI support, and experience sampling to gauge how AI use shapes study habits and help-seeking behavior over time. Similarly, for educators, we need to understand how teaching with AI tools alters lesson planning, formative assessment practices, or teacher professional identity. Research in this area would help identify best practices for human–AI collaboration in learning—clarifying when AI enhances understanding versus when it might hinder deeper engagement. It would also shed light on potential unintended consequences, such as overdependence on AI or the erosion of students' sense of academic agency, enabling the design of interventions to mitigate these risks.

Third, questions of effectiveness and scaling require much more evidence. A critical open question is whether agentic AI can approach the effectiveness of high-quality human tutoring at scale. Bloom's (1984) seminal work on the "two sigma problem" showed that one-on-one human tutoring produces learning gains about two standard deviations above conventional classroom instruction. Can AI tutors or coach agents achieve anything close to those results in real-world settings, and if so, under what conditions? So far, the evidence is mixed: some studies report large short-term boosts in homework completion or writing quality with AI support, while others find negligible or even negative effects once factors such as student motivation and oversight are considered. We lack comprehensive longitudinal studies following students over semesters or years to see if AI-assisted learning leads to better retention, transferable skills, or academic achievement (Giannakos et al. 2024). We also need experiments that vary how AI is integrated (e.g., as a supplement to teacher instruction versus a replacement for certain activities) to identify effective implementation models. Beyond academic outcomes, research should evaluate impacts on non-cognitive outcomes such as curiosity, resilience, and collaboration skills. Establishing a reliable evidence base on what agentic AI does—and does not—improve will help educational institutions make informed adoption decisions rooted in pedagogical benefit rather than hype.

Finally, the ethical and equity implications of agentic AI in education constitute a major research frontier. Educators and policymakers require guidance on how to navigate the value-laden choices that AI systems introduce. For instance, an AI teaching assistant might have to balance between pushing a struggling student with challenging tasks and ensuring the student does not become discouraged—a decision that reflects educational values. We need studies examining how teachers, students, and parents across different cultures and communities perceive such AI-driven decisions. Just as large cross-cultural surveys of AI decision-making in the context of self-driving cars (e.g., Awad et al. 2018) revealed divergent ethical priorities around the world, analogous research in education could show how stakeholders in different regions prioritize concerns such as data privacy versus personalization, or efficiency gains versus the preservation of human teacher roles. This line of inquiry would help develop context-sensitive ethical frameworks for AI in schools instead of one-size-fits-all policies. Furthermore, ensuring equity requires investigating whether AI tools inadvertently amplify biases or opportunity gaps. Research should audit AI educational systems for differential performance: Do language models provide lower-quality feedback on essays written in non-standard dialects or by non-native English speakers? Are predictive analytics in learning management systems accurately identifying at-risk students from underrepresented groups, or are they reflecting historical bias in the data? Ethical questions persist as well: AI cannot make the qualitative, compassionate judgments teachers routinely make—noticing a student seems hungry or remembering difficult circumstances at home—and we do not yet understand the impact of that absence. Such questions demand rigorous analysis. Addressing them will inform the design of governance, training, and monitoring approaches that keep AI use aligned with principles of fairness, transparency, and inclusion. In sum, research on the ethical, cultural, and equity dimensions of AI in education is essential to ensure that technological advances actually benefit all learners and uphold society's broader educational values.

# 6. Conclusion

AI use in education is no longer speculative. Learners and educators are adopting general-purpose and agentic AI tools at a pace that outstrips the ability of institutional strategies, infrastructure, and regulations to adapt. This mismatch creates real risk and real opportunity. Institutions that ignore or categorically resist AI drift away from how learning and academic work actually happen; institutions that engage with clear goals and disciplined designs can steer AI toward their educational purposes.

This chapter proposed the three-tension framework—Implementation Feasibility, Adaptation Speed, and Mission Alignment—as a practical lens for decision-making. Any serious AI initiative in education must ask whether it can be operated safely and sustainably, whether institutional learning and governance can keep pace with technical change, and whether proposed uses serve core values such as equity, transparency, and meaningful learning. These tensions interact: rapid deployment without guardrails strains feasibility and trust; strict protection without space for experimentation pushes activity into ungoverned use. Successful innovation requires coordinated attention to all three dimensions rather than isolated technical fixes.

From this perspective, responsible adoption follows a set of grounded principles. Start with concrete use cases and explicit success criteria instead of promising total transformation. Build monitoring and evaluation in from the beginning so deployments function as tested hypotheses, not one-off installations. Invest in human capacity: staff need time, tools, and recognition to develop AI literacy and to redesign tasks, assessment, and support. Keep consequential decisions—grading, progression, placement, welfare—under human authority, with agents providing evidence and recommendations rather than verdicts. Treat equity and ethics as design requirements: ensure access, protect privacy, require clarity about how systems work, and check for uneven benefits or harms across student groups.

The future trajectory of AI capabilities remains uncertain. Some analyses suggest current architectures may approach limits without new breakthroughs (Marcus 2025; Shojaee et al. 2025), while others expect steady gains and new applications. A prudent stance is to plan for both constraint and continued progress: assume present systems need human judgment and strong governance, and design policies flexible enough to absorb improvements without constant crisis revision. Historical experience warns against confident dismissal of emergent technologies: early predictions that online shopping would remain marginal (Time 1966), that the internet could not scale beyond a niche network (Metcalfe 1995), or that smartphones had “no chance” of meaningful market share (Anderson 2007) all underestimated how incremental advances and ecosystem effects turn experimental tools into basic infrastructure. Agentic AI could follow a similar path from scattered pilots to quiet backbone. Whether that shift benefits education depends less on technical inevitability and more on choices made now.

If teachers and education leaders do not make those choices, others will. For instance, commercial providers may pursue engagement or efficiency in ways that conflict with pedagogical values, while government regulators might impose one-size-fits-all mandates in reaction to lobbying pressure or high-profile failures. Likewise, students and staff may rely on consumer tools – sometimes concealed – in ways that lack transparency, alignment, safeguards, or support. In each case, institutions lose the chance to align AI with their mission. Active, intentional engagement offers a better route. Institutions can decide where AI should help, where it should be limited, and how it should be checked. This requires reimagining traditional practices, e.g., updating assessments so that students are rewarded for demonstrating reasoning and creativity, not just for producing declarative knowledge that an AI could generate for them. Without such adjustments, the same AI tools that can boost students’ productivity might also tempt some learners to bypass deeper learning, producing superficially improved work that masks a lack of understanding. Similarly, without proper professional development, teachers might either over-rely on AI (diminishing their personal engagement with students) or avoid it entirely out of uncertainty (missing the chance to improve instruction).

Ultimately, AI is emerging as another actor in the educational ecosystem, alongside teachers, students, and parents. Harnessing its potential upside requires that humans delegate some tasks and decisions to machines—but do so carefully, maintaining human oversight and accountability. The path forward, therefore, is one of active leadership and thoughtful experimentation. Rather than let the future of AI in education happen to us, we must take ownership of that future. By applying the three-tension framework—ensuring any AI initiative is feasible to implement, paced to learn and adapt, and aligned to core values—we increase the odds that this powerful technology will serve as a positive transformative force. The coming decade will test the resolve and agility of educational systems worldwide. Those that navigate the tensions with foresight and humanity will not only avoid the pitfalls of unchecked technological disruption, but also apply agentic AI to enrich learning, advance equity, and better prepare students for life in an AI-enhanced society.

# Acknowledgments

In preparing this chapter, we used various LLMs to assist with literature discovery, text editing, and proofreading. We reviewed and verified all outputs and are responsible for the content.

# References

AAUP Committee on Artificial Intelligence and Academic Professions. (2024). *Artificial intelligence and academic professions*. American Association of University Professors. https://www.aaup.org/reports-publications/aaup-policies-reports/topical-reports/artificial-intelligence-and-academic. Accessed 10 October 2025.

Agbo, F. J., Olivia, C., Oguibe, G., Sanusi, I. T., & Sani, G. (2025). Computing education using generative artificial intelligence tools: A systematic literature review. *Computers and Education Open, 9*, 100266. https://doi.org/10.1016/j.caeo.2025.100266

Al-Abdullatif, A. M., & Alsubaie, M. A. (2024). ChatGPT in learning: Assessing students' use intentions through the lens of perceived value and the influence of AI literacy. *Behavioral Sciences, 14*(9), 845. https://doi.org/10.3390/bs14090845

Alfarwan, A. (2025). Generative AI use in K–12 education: A systematic review. Frontiers in Education, 10, Article 1647573. https://doi.org/10.3389/feduc.2025.1647573

Ali, D., Fatemi, Y., Boskabadi, E., Nikfar, M., Ugwuoke, J., & Ali, H. (2024). ChatGPT in teaching and learning: A systematic review. *Education Sciences, 14*(6), 643. https://doi.org/10.3390/educsci14060643

American Association of University Professors. (1966). *Statement on government of colleges and universities*. American Association of University Professors. https://www.aaup.org/report/statement-government-colleges-and-universities. Accessed 10 October 2025.

Ammari, T., Chen, M., Zaman, S. M. M., & Garimella, K. (2025). How students (really) use ChatGPT: Uncovering experiences among undergraduate students. *arXiv*. https://arxiv.org/abs/2505.24126

Anderson, N. (2007, April 30). Ballmer: iPhone has "no chance" of gaining significant market share. *Ars Technica*. https://arstechnica.com/information-technology/2007/04/ballmer-says-iphone-has-no-chance-to-gain-significant-market-share. Accessed 7 October 2025.

Awad, E., Dsouza, S., Kim, R., Schulz, J., Henrich, J., Shariff, A., Bonnefon, J.-F., & Rahwan, I. (2018). The Moral Machine experiment. *Nature, 563*(7729), 59–64. https://doi.org/10.1038/s41586-018-0637-6

Baig, M. I., & Yadegaridehkordi, E. (2024). ChatGPT in the higher education: A systematic literature review and research challenges. *International Journal of Educational Research, 127*, 102411. https://doi.org/10.1016/j.ijer.2024.102411

Baytas, C., & Ruediger, D. (2025). *Making AI generative for higher education: Adoption and challenges among instructors and researchers*. Ithaka S+R. https://doi.org/10.18665/sr.322677

Bellan, R. (2025, October 6). Sam Altman says ChatGPT has hit 800 M weekly active users. *TechCrunch*. https://techcrunch.com/2025/10/06/sam-altman-says-chatgpt-has-hit-800m-weekly-active-users. Accessed 10 October 2025.

Becker, J., Rush, N., Barnes, B., & Rein, D. (2025). Measuring the impact of early-2025 AI on experienced open-source developer productivity (arXiv:2507.09089). *arXiv*. https://arxiv.org/abs/2507.09089

Behrens, J. T., DiCerbo, K. E., & Foltz, P. W. (2019). Assessment of complex performances in digital environments. *The ANNALS of the American Academy of Political and Social Science, 683*(1), 217–232. https://doi.org/10.1177/0002716219846850

Bent, D., Handa, K., Durmus, E., Tamkin, A., McCain, M., Ritchie, S., Donegan, R., Martinez, J., & Jones, J. (2025, August 27). Anthropic education report: How educators use Claude. Anthropic.

https://www.anthropic.com/news/anthropic-education-report-how-educators-use-claude. Accessed 10 October 2025.

Bloom, B. S. (1984). The 2 sigma problem: The search for methods of group instruction as effective as one-to-one tutoring. Educational Researcher, 13(6), 4–16. https://doi.org/10.3102/0013189X013006004

Bloom, D. E., Prettner, K., Saadaoui, J., & Veruete, M. (2024). Artificial intelligence and the skill premium (NBER Working Paper No. 32430). National Bureau of Economic Research. https://www.nber.org/papers/w32430

Brynjolfsson, E., Li, D., & Raymond, L. R. (2025). Generative AI at work. *The Quarterly Journal of Economics, 140*(2), 889–942. https://doi.org/10.1093/qje/qjae044

California Department of Education. (2021). *Ethnic studies model curriculum*. https://www.cde.ca.gov/ci/cr/cf/esmc.asp. Accessed 13 October 2025.

Center on Budget and Policy Priorities. (2024). *Expiration of federal K-12 emergency funds could pose challenges for states*. https://www.cbpp.org/research/state-budget-and-tax/expiration-of-federal-k-12-emergency-funds-could-pose-challenges-for. Accessed 13 October 2025.

Chapman, B. (2024, June 26). Turmoil surrounds LA's new AI student chatbot as tech firm furloughs staff just three months after launch. *The 74*. https://www.the74million.org/article/turmoil-surrounds-las-new-ai-student-chatbot-as-tech-firm-furloughs-staff-just-3-months-after-launch. Accessed 20 October 2025.

Chatterji, A., Cunningham, T., Deming, D. J., Hitzig, Z., Ong, C., Shan, C. Y., & Wadman, K. (2025, September). How people use ChatGPT (NBER Working Paper No. 34255). *National Bureau of Economic Research*. http://www.nber.org/papers/w34255. Accessed 29 October 2025.

*Children's Online Privacy Protection Act of 1998*. (1998). 15 U.S.C. §§ 6501–6506.

Chiu, T. K. F. (2025). Developing intelligent-TPACK (I-TPACK) framework from unpacking AI literacy and competency: Implementation strategies and future research direction. *Interactive Learning Environments, 33*(7), 4189–4192. https://doi.org/10.1080/10494820.2025.2545053

Choi, G. W., Kim, S. H., Lee, D., & Moon, J. (2024). Utilizing generative AI for instructional design: Exploring strengths, weaknesses, opportunities, and threats. *TechTrends, 68*, 832–844. https://doi.org/10.1007/s11528-024-00967-w

Chopra, A., Bhattacharya, S., Salvador, D., Paul, A., Wright, T., Garg, A., Ahmad, F., Schwarze, A. C., Raskar, R., & Balaprakash, P. (2025). The Iceberg Index: Measuring skills-centered exposure in the AI economy. *arXiv*. https://doi.org/10.48550/arXiv.2510.25137

Chui, M., Hazan, E., Roberts, R., Singla, A., Smaje, K., Sukharevsky, A., Yee, L., & Zemmel, R. (2023, June). *The economic potential of generative AI: The next productivity frontier*. McKinsey Global Institute.

Cohen, W. M., & Levinthal, D. A. (1990). Absorptive capacity: A new perspective on learning and innovation. *Administrative Science Quarterly, 35*(1), 128–152. https://doi.org/10.2307/2393553

Cook, J. (2024, July 16). OpenAI's 5 levels of "super AI" (AGI to outperform human capability). Forbes.https://www.forbes.com/sites/jodiecook/2024/07/16/openais-5-levels-of-super-ai-agi-to-outperform-human-capability. Accessed 11 October 2025.

CoSN, & CGCS. (2024). *AI readiness rubric for K-12 school districts*. Consortium for School Networking. https://www.cosn.org. Accessed 29 October 2025.

Deng, R., Jiang, M., Yu, X., Lu, Y., & Liu, S. (2025). Does ChatGPT enhance student learning? A systematic review and meta-analysis of experimental studies. *Computers & Education, 227*, 105224. https://doi.org/10.1016/j.compedu.2024.105224

Dell'Acqua, F., McFowland, E., III, Mollick, E., Lifshitz-Assaf, H., Kellogg, K. C., Rajendran, S., Krayer, L., Candelon, F., & Lakhani, K. R. (2023). Navigating the jagged technological frontier: Field experimental evidence of the effects of AI on knowledge worker productivity and quality (Working Paper No. 24-013). *Harvard Business School*. https://doi.org/10.2139/ssrn.4573321

DiCerbo, K. E., & Behrens, J. T. (2014). *Impacts of the digital ocean on education*. Pearson.

Diliberti, M. K., Lake, R. J., & Weiner, S. R. (2025). *More districts are training teachers on artificial intelligence: Findings from the American School District Panel* (Report No. RR-A956-31). RAND Corporation. https://www.rand.org/pubs/research_reports/RRA956-31.html. Accessed 29 October 2025.

Diliberti, M. K., Schwartz, H. L., Doan, S., Shapiro, A., Rainey, L. R., & Lake, R. J. (2024). *Using artificial intelligence tools in K–12 classrooms* (Research Report No. RR-A956-21). RAND Corporation. https://doi.org/10.7249/RRA956-21

Doss, C. J., Bozick, R., Schwartz, H. L., Chu, L., Rainey, L. R., Woo, A., Reich, J., & Dukes, J. (2025). *AI use in schools is quickly increasing but guidance lags behind: Findings from the RAND Survey Panels* (RR-A4180-1). RAND Corporation. https://www.rand.org/t/RRA4180-1. Accessed 23 October 2025.

EDUCAUSE. (2024). *AI readiness framework for higher education*. https://www.educause.edu. Accessed 23 October 2025.

EDUCAUSE. (2025). *AI ethics framework for higher education*. https://www.educause.edu. Accessed 23 October 2025.

Eggerton, J. (2019). FCC: Cloud equivalents are eligible for E-Rate. *Next TV*. https://www.nexttv.com/news/fcc-cloud-equivalents-are-eligible-for-e-rate. Accessed 23 October 2025.

Eloundou, T., Manning, S., Mishkin, P., & Rock, D. (2024). GPTs are GPTs: Labor market impact potential of LLMs. *Science, 384*, 1306–1308. https://doi.org/10.1126/science.adj0998

Ertmer, P. A. (1999). Addressing first- and second-order barriers to change: Strategies for technology integration. *Educational Technology Research and Development, 47*(4), 47–61. https://doi.org/10.1007/BF02299597

*Family Educational Rights and Privacy Act of 1974*. (1974). 20 U.S.C. § 1232g.

Federal Communications Commission, Wireline Competition Bureau. (2019). *Eligible services list for funding year 2020*. https://docs.fcc.gov/public/attachments/DA-18-1173A1.pdf. Accessed 23 October 2025.

Figueroa de la Fuente, M., & Farhadian, G. (2025). Systematic review of the impact of artificial intelligence in higher education. *Journal of Teaching and Learning, 19*(4), 72–96. https://doi.org/10.22329/jtl.v19i4.9849

Fittes, E. K. (2025, August 28). K-12 sales cycle: How long do districts need to make smart purchasing decisions? *EdWeek Market Brief*. https://marketbrief.edweek.org/sales-marketing/k-12-sales-cycle-how-long-do-districts-need-to-make-smart-purchasing-decisions/2025/08. Accessed 23 October 2025.

Freeman, S., Eddy, S. L., McDonough, M., Smith, M. K., Okoroafor, N., Jordt, H., & Wenderoth, M. P. (2014). Active learning increases student performance in science, engineering, and mathematics. *Proceedings of the National Academy of Sciences, 111*(23), 8410–8415. https://doi.org/10.1073/pnas.1319030111

Gallup, & Walton Family Foundation. (2025). *Teaching for tomorrow: Unlocking six weeks a year with AI*. https://www.gallup.com/analytics/659819/k-12-teacher-research.aspx. Accessed 23 October 2025.

García, E., & Weiss, E. (2022). *Raising pay in public K–12 schools is critical to solving staffing shortages*. Economic Policy Institute. https://www.epi.org/publication/solving-k-12-staffing-shortages. Accessed 23 October 2025.

Gartner. (2025, June 25). *Gartner predicts over 40% of agentic AI projects will be cancelled by end of 2027*. [Press release]. https://www.gartner.com/en/newsroom/press-releases/2025-06-25-gartner-predicts-over-40-percent-of-agentic-ai-projects-will-be-canceled-by-end-of-2027. Accessed 23 October 2025.

Gasser, U., & Schmitt, C. (2023). *Regulatory sandboxes in artificial intelligence*. Berkman Klein Center for Internet & Society Research Publication Series.

Giannakos, M., Azevedo, R., Brusilovsky, P., Cukurova, M., Dimitriadis, Y., Hernandez-Leo, D., Järvelä, S., Mavrikis, M., & Rienties, B. (2024). The promise and challenges of generative AI in education. *Behaviour & Information Technology*. https://doi.org/10.1080/0144929X.2024.2394886

Gillespie, N., Lockey, S., Ward, T., Macdade, A., & Hassed, G. (2025). Trust, attitudes and use of artificial intelligence: A global study 2025. University of Melbourne and KPMG International. https://doi.org/10.26188/28822919

Gimbel, M., Kinder, M., Kendall, J., & Lee, M. (2025, October 1). *Evaluating the impact of AI on the labour market: Current state of affairs*. The Budget Lab at Yale University. https://budgetlab.yale.edu/research/evaluating-impact-ai-labor-market-current-state-affairs. Accessed 21 October 2025.

Google DeepMind. (2025, April 2). *Taking a responsible path to AGI*. https://deepmind.google/discover/blog/taking-a-responsible-path-to-agi. Accessed 23 October 2025.

Handa, K., Bent, D., Tamkin, A., McCain, M., Durmus, E., Stern, M., Schiraldi, M., Huang, S., Ritchie, S., Syverud, S., Jagadish, K., Vo, M., Bell, M., & Ganguli, D. (2025, April 8). Anthropic education report: How university students use Claude. Anthropic. https://www.anthropic.com/news/anthropic-education-report-how-university-students-use-claude. Accessed 23 October 2025.

Hardman, P. (2025). The changing relationship between education and tech: What Google DeepMind's AI for Learning Forum signals about the next era of ed-tech. Dr Phil's Newsletter. https://philippahardman.substack.com. Accessed 9 November 2025.

Hendriksen, M., & Lai, E. (2025). *Asking to learn: What student queries to generative AI reveal about cognitive engagement*. Pearson.

Hendrycks, D., Song, D., Szegedy, C., Lee, H., Gal, Y., Brynjolfsson, E., et al. (2025, October 23). A definition of AGI (arXiv:2510.18212v2) [Preprint]. *arXiv*. https://arxiv.org/abs/2510.18212

Huang, K.-H., Prabhakar, A., Thorat, O., Agarwal, D., Choubey, P. K., Mao, Y., Savarese, S., Xiong, C., & Wu, C.-S. (2025, May 24). CRMArena-Pro: Holistic assessment of LLM agents across diverse business scenarios and interactions [Preprint]. *arXiv*. https://arxiv.org/abs/2505.18878

Imagine Learning. (2025). *Curriculum-informed AI™: The next era of AI in education*. Imagine Learning. https://www.imaginelearning.com/wp-content/uploads/2025/04/1620957897_Curriculum_Informed_AI_White_Paper_DIGITAL.pdf. Accessed 23 October 2025.

ISTE. (2025). *ISTE standards and essential conditions for AI implementation*. International Society for Technology in Education. https://www.iste.org. Accessed 25 October 2025.

Jensen, L. X., Buhl, A., Sharma, A., & Bearman, M. (2025). Generative AI and higher education: A review of claims from the first months of ChatGPT. *Higher Education, 89*, 1145–1161. https://doi.org/10.1007/s10734-024-01265-3

K–12 Dive. (2024, September 30). *With ESSER expiration, COVID-19 spending prepares for finale*. https://www.k12dive.com/news/school-funding-esser-American-Rescue-Plan-covid-19-coronavirus/728352. Accessed 25 October 2025.

Kergroach, S., & Héritier, J. (2025, June 25). *Emerging divides in the transition to artificial intelligence* (OECD Regional Development Papers No. 147). OECD Publishing. https://doi.org/10.1787/7376c776-en

Koenecke, A., Nam, A., Lake, E., Nudell, J., Quartey, M., Mengesha, Z., Toups, C., Rickford, J. R., Jurafsky, D., & Goel, S. (2020). Racial disparities in automated speech recognition. Proceedings of the National Academy of Sciences, 117(14), 7684–7689. https://doi.org/10.1073/pnas.1915768117

Kozan, K., Hur, J., Kim, I., & Barrett, A. (2025). Instructional designers' integration of generative artificial intelligence into their professional practice. *Education Sciences, 15*(9), 1133. https://doi.org/10.3390/educsci15091133

Krishna Nk, R., Anupama, R. S., & Shivaraj, K. (2025). Artificial intelligence in radiology: Augmentation, not replacement. *Cureus, 17*(6), e86247. https://doi.org/10.7759/cureus.86247

Liang, W., Yuksekgonul, M., Mao, Y., Wu, E., & Zou, J. (2023). GPT detectors are biased against non-native English writers. Patterns, 4(8), 100779. https://doi.org/10.1016/j.patter.2023.100779

Lee, D., Arnold, M., Srivastava, A., Plastow, K., Strelan, P., Ploeckl, F., Lekkas, D., & Palmer, E. (2024). The impact of generative AI on higher education learning and teaching: A study of educators' perspectives. *Computers and Education: Artificial Intelligence, 6*, 100221. https://doi.org/10.1016/j.caeai.2024.100221

Lin, X., & Tan, H. (2025). A systematic review of generative AI in K–12: Mapping goals, activities, roles, and outcomes via the 3P model. *Systems, 13*(10), 840. https://doi.org/10.3390/systems13100840

Lincoln Institute of Land Policy. (2022). Public schools and the property tax: A comparison of education funding models in three US states. *Land Lines*. https://www.lincolninst.edu/publications/articles/2022-04-public-schools-property-tax-comparison-education-models. Accessed 25 October 2025.

Liu, R., Zenke, C., Liu, C., Holmes, A., Thornton, P., & Malan, D. J. (2024). Teaching CS50 with AI: Leveraging generative artificial intelligence in computer science education. In *Proceedings of the 55th ACM Technical Symposium on Computer Science Education V. 1 (SIGCSE 2024)* (pp. 123–129). Association for Computing Machinery. https://doi.org/10.1145/3626252.3630938

Luo, T., Muljana, P. S., Ren, X., & Young, D. (2025). Exploring instructional designers' utilisation and perspectives on generative AI tools: A mixed methods study. *Educational Technology Research and Development, 73*(2), 741–766. https://doi.org/10.1007/s11423-024-10437-y

Łodzikowski, K., Foltz, P. W., & Behrens, J. T. (2024). Generative AI and its educational implications. In D. Kourkoulou, A.-O. Tzirides, B. Cope, & M. Kalantzis (Eds.), *Trust and inclusion in AI-mediated education: Where human learning meets learning machines* (pp. 35–57). Springer. https://doi.org/10.1007/978-3-031-64487-0_2

Mai, D. T. T., Da, C. V., & Hanh, N. V. (2024). The use of ChatGPT in teaching and learning: A systematic review through SWOT analysis approach. *Frontiers in Education, 9*, 1328769. https://doi.org/10.3389/feduc.2024.1328769

Mäkelä, M., & Stephany, F. (2024). Artificial intelligence and labour market inequality: Skill premiums, displacement, and adaptation. *Research Policy, 53*(4), 104964.

Marcus, G. (2025, October 18). Game over. AGI is not imminent, and LLMs are not the royal road to getting there. *Marcus on AI*. https://garymarcus.substack.com/p/the-last-few-months-have-been-devastating. Accessed 2 November 2025.

Martin, B. (2025, September 2). *Georgia Tech's Jill Watson outperforms ChatGPT in real classrooms*. Georgia Tech Research. https://research.gatech.edu/georgia-techs-jill-watson-outperforms-chatgpt-real-classrooms

Mazeika, M., Gatti, A., Menghini, C., Sehwag, U. M., Singhal, S., Orlovskiy, Y., Basart, S., Sharma, M., Peskoff, D., Lau, E., Lim, J., Carroll, L., Blair, A., Sivakumar, V., Basu, S., Kenstler, B., Ma, Y., Michael, J., Li, X., … Hendrycks, D. (2025). Remote labor index: Measuring AI automation of remote work [Preprint]. arXiv. https://doi.org/10.48550/arXiv.2510.26787

McGehee, N. (2024). *Breaking barriers: A meta-analysis of educator acceptance of AI technology in education*. Michigan Virtual Learning Research Institute. https://michiganvirtual.org/research/publications/breaking-barriers-a-meta-analysis-of-educator-acceptance-of-ai-technology-in-education. Accessed 25 October 2025.

McKinsey & Company. (2025a). *Seizing the agentic AI advantage*. McKinsey & Company.

McKinsey & Company. (2025b). *The state of AI: Global survey*. McKinsey & Company.

Metcalfe, R. (1995, December 4). From the ether: Predicting the internet's catastrophic collapse and ghost sites galore in 1996. *InfoWorld*.

http://www.infoworld.com/cgi-bin/displayNew.pl?/metcalfe/bm120495.htm. Accessed 25 October 2025.

Miller, M. T., & Nadler, D. P. (2019). Faculty senates and college presidents: Perspectives on collaboration. *Journal of Research on the College President*. https://collegepresidentresearch.uark.edu/2019/12/faculty-senates-and-college-presidents-perspectives-on-collaboration. Accessed 25 October 2025.

Mishra, P., & Koehler, M. J. (2006). Technological pedagogical content knowledge: A framework for teacher knowledge. *Teachers College Record, 108*(6), 1017–1054.

Mousa, D. (2025, September 25). AI isn't replacing radiologists. *Works in Progress*. https://www.worksinprogress.news/p/why-ai-isnt-replacing-radiologists. Accessed 25 October 2025.

Mukherjee, S. (2017, March 27). A.I. versus M.D.: What happens when diagnosis is automated? *The New Yorker*. https://www.newyorker.com/magazine/2017/04/03/ai-versus-md. Accessed 25 October 2025.

National Center for Education Statistics. (2024a). *English learners in public schools*. *Condition of Education*. US Department of Education, Institute of Education Sciences. https://nces.ed.gov/programs/coe/indicator/cgf. Accessed 25 October 2025.

National Center for Education Statistics. (2024b). *Racial/ethnic enrollment in public schools*. *Condition of Education*. US Department of Education. https://nces.ed.gov/programs/coe/indicator/cge. Accessed 25 October 2025.

National Center for Education Statistics. (2024c). *Students with disabilities*. *Condition of Education*. US Department of Education. https://nces.ed.gov/programs/coe/indicator/cgg. Accessed 25 October 2025.

Noy, S., & Zhang, W. (2023). Experimental evidence on the productivity effects of generative artificial intelligence. *Science, 381*(6654), 187–192. https://doi.org/10.1126/science.adh2586

OECD. (2024). *Explanatory memorandum on the updated OECD definition of an AI system*. OECD Publishing. https://doi.org/10.1787/623da898-en

OECD. (2025a). *Introducing the OECD AI capability indicators*. OECD Publishing. https://doi.org/10.1787/be745f04-en

OECD. (2025b). *Results from TALIS 2024: United States (TALIS 2024 country note)*. OECD Publishing.

OpenAI. (2018, April 9). OpenAI charter. https://openai.com/charter. Accessed 25 October 2025.

OpenAI. (2023, March). GPT-4. OpenAI Blog. https://openai.com/index/gpt-4-research/. Accessed 25 October 2025.

OpenAI. (2024, September). Introducing o1. OpenAI Blog. https://openai.com/blog. Accessed 25 October 2025.

Pan, Y., Kong, D., Zhou, S., Cui, C., Leng, Y., Jiang, B., Liu, H., Shang, Y., Zhou, S., Wu, T., & Wu, Z. (2024, July 16). WebCanvas: Benchmarking web agents in online environments (arXiv:2406.12373) [Preprint]. *arXiv*. https://doi.org/10.48550/arXiv.2406.12373.

Pearson. (2019). *Aida Calculus* [Mobile app]. Apple App Store. https://apps.apple.com. Accessed 22 October 2025.

Pearson. (2025a). *AI-powered study tools: The Pearson difference*. Pearson. https://www.pearson.com/content/dam/global-store/en-us/resources/Pearson-Difference-AI-Feature-Digital.pdf. Accessed 22 October 2025.

Pearson. (2025b). *Higher ed student AI survey report*. Pearson.https://plc.pearson.com/sites/pearson-corp/files/2025-05/student-ai-tracker-spring-25-short-version-for-external-sharing.pdf. Accessed 22 October 2025.

Peng, S., Kalliamvakou, E., Cihon, P., & Demirer, M. (2023). The impact of AI on developer productivity: Evidence from GitHub Copilot (arXiv:2302.06590). *arXiv*. https://arxiv.org/abs/2302.06590

PwC. (2025, May). AI agent survey. https://www.pwc.com/us/en/tech-effect/ai-analytics/ai-agent-survey.html

Rogers, E. M. (2003). *Diffusion of innovations* (5th ed.). Free Press.

Russell, S. J., & Norvig, P. (2020). *Artificial intelligence: A modern approach* (4th ed.). Pearson.

Schoorman, D., & Gatens, R. (2025). Understanding Florida's HB7: A policy of intimidation by confusion. *Journal of Education Policy*. Advance online publication. https://doi.org/10.1177/08959048241235405

Shojaee, P., Mirzadeh, I., Alizadeh, K., Horton, M., Bengio, S., & Farajtabar, M. (2025). The illusion of thinking: Understanding the strengths and limitations of reasoning models via the lens of problem complexity. *arXiv*. https://doi.org/10.48550/arXiv.2506.06941

TeachAI. (2025). *Guidance for generative AI in education*. https://www.teachai.org. Accessed 22 October 2025.

Texas Legislature. (2021). *House Bill 3979: Relating to the social studies curriculum in public schools; and Senate Bill 3*. https://legiscan.com/TX/supplement/HB3979/id/197235. Accessed 22 October 2025.

Time. (1966, February 25). The futurists: Looking toward A.D. 2000. *Time*. https://content.time.com/time/subscriber/article/0%2C33009%2C835128-5%2C00.html. Accessed 22 October 2025.

UNESCO. (2021). *AI and education: Guidance for policy-makers*. UNESCO Publishing.

UNESCO. (2025, September 2). *UNESCO survey: Two-thirds of higher education institutions have or are developing guidance on AI use*. https://www.unesco.org/en/articles/unesco-survey-two-thirds-higher-education-institutions-have-or-are-developing-guidance-ai-use. Accessed 22 October 2025.

US Department of Education. (2019). *Supplement not supplant under Title I, Part A of the Elementary and Secondary Education Act of 1965, as amended by the Every Student Succeeds Act: Non-regulatory informational document*. https://www.ed.gov/sites/ed/files/2020/02/snsfinalguidance06192019.pdf. Accessed 22 October 2025.

US Department of Justice, Civil Rights Division. (2024, April 24). Nondiscrimination on the basis of disability; Accessibility of web information and services of state and local government entities [Final rule]. *Federal Register, 89*, 31320–31568.

US Government Accountability Office. (2020). *K-12 education: School districts frequently identified multiple building systems needing updates or replacement* (GAO-20-494). https://www.gao.gov/products/gao-20-494. Accessed 22 October 2025.

Wang, L., Ma, C., Feng, X., Zhang, Z., Yang, H., Zhang, J., Chen, Z.-Y., Tang, J., Chen, X., Lin, Y., Zhao, W. X., Wei, Z., & Wen, J.-R. (2025). A survey on large language model-based autonomous agents. *Frontiers of Computer Science*. Advance online publication. https://doi.org/10.1007/s11704-024-40231-1

Wang, Y., Wei, Z., Wijaya, T. T., Cao, Y., & Ning, Y. (2025). Awareness, acceptance, and adoption of Gen-AI by K-12 mathematics teachers: An empirical study integrating TAM and TPB. *BMC Psychology, 13*, 478. https://doi.org/10.1186/s40359-025-01865-4

Weiner, S., Lake, R., & Rosner, J. (2024, October). *AI is evolving, but teacher prep is lagging: A first look at teacher preparation program responses to AI*. Center on Reinventing Public Education & AACTE. https://crpe.org/wp-content/uploads/Teacher-Prep-AI-2024.pdf. Accessed 22 October 2025.

Xie, T., Zhang, D., Chen, J., Li, X., Zhao, S., Cao, R., Hua, T. J., Cheng, Z., Shin, D., Lei, F., Liu, Y., Xu, Y., Zhou, S., Savarese, S., Xiong, C., Zhong, V., & Yu, T. (2024). OSWorld: Benchmarking multimodal agents for open-ended tasks in real computer environments [Preprint]. *arXiv*. https://doi.org/10.48550/arXiv.2404.07972

Zhou, S., Xu, F. F., Zhu, H., Zhou, X., Lo, R., Sridhar, A., Cheng, X., Ou, T., Bisk, Y., Fried, D., Alon, U., & Neubig, G. (2024). WebArena: A realistic web environment for building autonomous agents [Preprint]. *arXiv*. https://doi.org/10.48550/arXiv.2307.13854